\documentclass[usenatbib]{mn2e}

\usepackage{graphicx}   
\usepackage{amsmath}    
\usepackage{amssymb}    

\title[First $z\sim6$ quasars in KiDS and VIKING]{First discoveries of
  \boldmath{$z\sim6$} quasars with the Kilo Degree Survey and VISTA
  Kilo-Degree Infrared Galaxy survey\thanks{Based on observations collected at
    the European Southern Observatory, Chile, programmes 177.A-3016,
    179.A-2004, 089.A-0290, 089.A-0596, 090.A-0383, 090.A-0642 and
    091.A-0421.}}  
\author[B. P. Venemans et al.]{B.~P.~Venemans,$^{1}$\thanks{E-mail:
    venemans@mpia.de} G.~A.~Verdoes Kleijn,$^{2}$ J.~Mwebaze,$^{2}$
  E.~A.~Valentijn,$^{2}$ \newauthor 
  E.~Ba\~nados,$^{1}$ R.~Decarli,$^{1}$
  J.~T.~A.~de Jong,$^{3}$ J.~R.~Findlay,$^{4}$ K.~H.~Kuijken,$^{3}$
\newauthor
  F.~La Barbera,$^{5}$ J.~P.~McFarland,$^{2}$ 
  R.~G.~McMahon,$^{6,7}$ N.~Napolitano,$^{5}$ 
\newauthor
  G.~Sikkema$^{2}$ and W.~J.~Sutherland$^{8}$
  \\
  $^{1}$Max-Planck Institute for Astronomy, K\"onigstuhl 17, D-69117
  Heidelberg, Germany \\
  $^{2}$Kapteyn Astronomical Institute, University of Groningen, PO Box 800, NL-9700 AV Groningen, The
  Netherlands \\
  $^{3}$ Leiden Observatory, Leiden University, P.O. Box 9513, NL-2300 RA Leiden, The Netherlands \\
  $^{4}$ Department of Physics and Astronomy, University of Wyoming,
  Laramie, WY 82071, USA \\
  $^{5}$ Astronomical Observatory of Capodimonte - INAF, via Moiariello 16, I-80131, Napoli, Italy \\
  $^{6}$ Institute of Astronomy, University of Cambridge, Madingley Road,
  Cambridge CB3 0HA, UK \\
  $^{7}$ Kavli Institute for Cosmology, University of Cambridge, Madingley
  Road, Cambridge CB3 0HA, UK \\
  $^{8}$ Astronomy Unit, School of Mathematical Sciences, Queen Mary,
  University of London, London, E1 4NS, UK \\
}

\date{Accepted 2015 July 31. Received 2015 July 31; in original form 2015 June 30}

\pubyear{2015}

\begin{document}
\label{firstpage}
\pagerange{\pageref{firstpage}--\pageref{lastpage}} 
\maketitle

\begin{abstract}
  We present the results of our first year of quasar search in the
  ongoing ESO public Kilo Degree Survey (KiDS) and VISTA Kilo-Degree
  Infrared Galaxy (VIKING) surveys. These surveys are among the deeper
  wide-field surveys that can be used to uncover large numbers of
  $z\sim6$ quasars. This allows us to probe a more common population
  of $z\sim6$ quasars that is fainter than the well-studied quasars
  from the main Sloan Digital Sky Survey. From this first set
  of combined survey catalogues covering $\sim$250\,deg$^2$ we
  selected point sources down to $Z_\mathrm{AB}=22$ that had a very
  red $i-Z$ ($i-Z>2.2$) colour. After follow-up imaging and
  spectroscopy, we discovered four new quasars in the redshift range
  $5.8<z<6.0$. The absolute magnitudes at a rest-frame wavelength of
  1450\,\AA\ are between $-26.6 < M_{1450} < -24.4$, confirming that
  we can find quasars fainter than $M^*$, which at $z=6$ has been
  estimated to be between $M^*=-25.1$ and $M^*=-27.6$. The discovery
  of four quasars in 250\,deg$^2$ of survey data is consistent with
  predictions based on the $z\sim6$ quasar luminosity function. We
  discuss various ways to push the candidate selection to fainter
  magnitudes and we expect to find about 30 new quasars down to an
  absolute magnitude of $M_{1450}=-24$. Studying this homogeneously
  selected faint quasar population will be important to gain insight
  into the onset of the co-evolution of the black holes and their
  stellar hosts.
\end{abstract}


\section{Introduction}

Supermassive black holes (SMBHs) are inferred to reside at the nuclei of most
(if not all) massive galaxies. In present day galaxies, tight correlations are
observed between the black hole mass and global host parameters over at
least three orders of magnitude in black hole mass \citep[see e.g][for a
  recent review]{kor13}. Going back in time, the global history of
star formation and of radiative mass accretion on to black holes
display similar behaviour as a function of time, peaking at redshifts
$\sim 2-3$ (\citealt{hec04}; for a review see \citealt{mad14} and
references therein). Determining the physical processes that govern
the symbiotic relationship between SMBHs
and their host galaxies is an outstanding challenge in astrophysics.
Studying the SMBHs and stellar hosts of high-redshift ($z>5.7$)
quasars is one of the most direct ways to obtain insight on the onset
of this relation in the first Gyr after the big bang. They provide
fundamental constraints on the formation and growth of the first
SMBHs, on early star formation and on
chemical enrichment in galaxy hosts and the surrounding intergalactic
matter \citep[e.g.][]{fan06b,jia07,der11,car13}.

Thanks to large imaging surveys of moderate depth, it has become
feasible to identify the rare bright quasars photometrically at high
redshifts. As an illustration, a luminous quasar at redshift 6 can be
detected photometrically in the $z$ band within two minutes on a 2.5 m
telescope. In the past decade about 70 bright quasars at $z>5.7$ have
been discovered
\citep[e.g.][]{fan06b,jia09,wil10a,mor11,ban14,ven15a} using a
variety of different multiwavelength surveys such as the Sloan
Digital Sky Survey \citep[SDSS;][]{aba09}, the Canada-France High-$z$
Quasar Survey \citep[CFHQS;][]{wil10a}, the UK Infrared Telescope
Infrared Deep Sky Survey \citep[UKIDSS;][]{law07} and Panoramic Survey
Telescope \& Rapid Response System 1
\citep[Pan-STARRS1;][]{mor12,ban14}.

These quasars appear to have SMBH masses equivalent to today's most
massive galaxies with high-accretion rates, close to Eddington. The
masses have been inferred from quasar emission line width and
continuum luminosity extrapolating empirical relationships established
in low redshift active galactic nuclei (AGN) through multi-epoch
reverberation mapping. For example, \citet{der11} obtain for a
population of $z\sim6$ quasars a mean black hole mass of
log($M_\mathrm{BH}$/M$_{\sun}$)\,$\approx9.06$, typical bolometric
luminosities $>10^{47}$ erg s$^{-1}$ and Eddington ratios $L_{\rm
  bol}$/$L_{\rm Edd}$\,$\approx 0.6$. The growth time for such black
holes and Eddington ratios if starting from $10^2$\,M$_{\sun}$ seeds
is comparable to the age of the Universe at $z\sim6$
($\sim$0.9\,Gyr). The quasar emission lines also suggest
well-developed metallicities. The relative abundance of Fe and
$\alpha$ elements, inferred from Fe\,{\sevensize II} and
Mg\,{\sevensize II} emission lines, appears similar to quasars at
redshift $\sim 4$ \citep[e.g.][]{jia07,der11}.
     
With the advent of ALMA it is now possible to obtain constraints on
host dynamical masses and on star formation by observing the cool
molecular gas of the hosts. The quasar hosts appear smaller or at most
roughly consistent with the present-day relation between host mass and
SMBH mass \citep[e.g.][]{wal04,wal09b,wan10,wan13,wil15}. These
dynamical masses are based on line kinematics. Direct detection of the
stellar hosts requires higher spatial resolution and has to await the
{\it James Webb Space Telescope} and the next generation ground-based
30\,m-class telescopes (see e.g.\ \citealt{mec12}). The star formation
rates measured in the quasar host galaxies are tens to hundred solar
masses per year \citep[e.g.][]{wan08b,wan13,wal09b,ven12}.

The results listed above are mostly based on the brightest (`tip of
the iceberg') of the quasar population. The quasars have been
selected close to the detection limit of flux-limited surveys and
represent the bright end of a more common fainter population. They
were detectable thanks to the combination of high SMBH mass with high
accretion rates. The few notable exceptions are the
$\sim$$10^8$\,M$_{\sun}$ SMBHs found by \citet{wil10b} and
\citet{kas15}.

Given the steepness of the quasar luminosity function (see
e.g.\ Fig.~\ref{f:lumfunct}; \citealt{wil10a,kas15}), a 2.5\,mag (a
factor 10 in continuum flux density) deeper photometric selection
would open the door to a quasar population which is expected to be an
order of magnitude more abundant. For a given accretion rate
(black hole mass) this population would probe active galaxies with
an order of magnitude lower black hole mass (accretion rate).

The abundance of SMBHs at lower masses and accretion rates is a
valuable constraint for the various alternative theories for the
formation and early evolution of SMBHs and their hosts. These include
the remnants of Population III stars, yielding black hole seeds of
the order of $10^2$\,M$_{\sun}$, direct collapse of primordial dense gas in
protogalaxies leading to masses of the order of $10^{4-5}$\,M$_{\sun}$ and
run-away dynamical instabilities in dense stellar systems which could
result in initial black hole masses of the order of $10^{2-3}$\,M$_{\sun}$
\citep[see e.g.][for a review]{vol10}. The theories predict average
mass densities for SMBHs seeds at $z\gg6$ that differ by up to an
order of magnitude. Hydrodynamical simulations are being used to
explore in detail these different formation channels. They are now
able to predict the relative abundance of bright and faint quasars
\citep[e.g.][]{cos14,kat15,lat15}. Thus establishing the quasar
demography (accretion rates, star formation rates, host masses) for
$\sim$ $10^8$\,M$_{\sun}$ SMBHs provides a direct discriminant on the
simulated formation scenarios.

The demography of quasars with $\sim$$10^8$\,M$_{\sun}$ SMBH masses is
also important for a fair comparison between the SMBH--host galaxy
relations at high redshift and today. For today's SMBH--host
relations, selecting black holes which are 10 times lower in mass
corresponds to probing stellar hosts which are 10--100 times more
abundant, are $\sim$2 mag fainter and have $\sim$35\% lower
stellar velocity dispersions. At black hole masses of
$10^8$\,M$_{\sun}$ we are also not hampered anymore by the potentially
severe selection bias in comparing AGN-selected samples at high
redshift to host-selected samples at low redshift. It could be that
current high-redshift samples are overwhelmed by exceptionally massive
$10^9$\,M$_{\sun}$ BHs in the more common smaller hosts compared to
the much rarer massive galaxies with equally massive black holes
\citep[e.g.][]{wil05b,lau07}.

For these reasons we are building up a homogeneous sample of faint
quasars at $z\sim6$ by combining ESO's public surveys the Kilo Degree
Survey (KiDS) and the VISTA Kilo-Degree Infrared Galaxy (VIKING)
survey which cover the same area \citep{arn07,dej13a}. The nine-band $u$
through $K_s$ photometry from the combined surveys goes up to $\sim$2
mag deeper than SDSS, UKIDSS and Pan-STARRS1. Here we report on the
high-$z$ quasar harvest from the first year of operations: four quasars
in range $5.8<z<6.0$. The results suggest we can expect to build up a
homogeneous sample of the order of 30 faint quasars at $z\sim 6$.

This paper is organized as follows. In Section~\ref{s:kidssurvey}, we
present the survey data from KiDS and VIKING we made use of and their
calibration. The selection of quasar candidates is detailed in
Section~\ref{s:selection}, followed by a description of the imaging
and spectroscopic follow-up in Sections~\ref{s:imaging} \&
\ref{s:spectroscopy}. The newly discovered quasars are presented in
Section~\ref{s:newqsos} and we conclude with a discussion and outlook in
Section~\ref{s:summary}.

In this paper, we adopt a cosmology with $\Omega_{\mathrm M}=0.28$,
$\Omega_{\Lambda}=0.72$ and $H_0=70$ km\,s$^{-1}$\,Mpc$^{-1}$
\citep{kom11}. All magnitudes are given in the AB system.

\section{KiDS and VIKING surveys and data}

\label{s:kidssurvey}

KiDS and VIKING are sister surveys that cover the same 1500\,deg$^2$
on the sky \citep{arn07,dej13a}. The survey areas are divided over two
main strips located in the Northern and Southern Galactic Cap (NGP and
SGP). The two strips, roughly equal in area, span $\sim$10$^\circ$ in
declination centred on $(\alpha,\delta)$=(12.5h,$0^\circ$) and
$(\alpha,\delta)$=(0.8h,$-30^\circ$). A smaller strip is centred on
the GAMA09 region with $(\alpha,\delta)$=(9h,$0^\circ$). KiDS covers
the optical Sloan $u$, $g$, $r$ and $i$ bands using the OmegaCAM
wide-field imager \citep{kui11} on the VLT Survey Telescope (VST) at
ESO's Paranal Observatory. VIKING covers the near-infrared $Z$
(hereafter $Z_V$), $Y$, $J$, $H$ and $K_s$ using the VISTA InfraRed
CAMera \citep[VIRCAM;][]{dal06} at the VISTA telescope \citep{eme06}
also at Paranal. VIKING and KiDS started regular survey operations on
2010 February 15 and 2011 October 15, respectively. The exposure times
and depth of the filters used in these two surveys are summarized in
Table~\ref{tab:exptimes}.

In this paper we focus on a set of 267 KiDS $i$ band slightly
overlapping pointings, each with an $\sim$1.15\,deg$^2$ field of
view. They were observed between the nights of 2011 September 2 and
2012 June 7, mostly during bright Moon. Due to slightly different
areal scheduling of KiDS and VIKING, the total area with $i$, $Z_V$
and $Y$ band coverage was 254.3\,deg$^2$ with 96.1\,deg$^2$ covered in
the NGP, 93.2\,deg$^2$ in the SGP and 65.0\,deg$^2$ was covered in
GAMA09. The search for high redshift quasars in this early phase of
KiDS operations was approached as a pilot project for the KiDS QSO
team. Its aim was to verify the observational quality (in terms of
photometry, astrometry, artefacts, data foibles) for quasar candidate
selection and the turn-around time between observation and final
confirmation via follow-up spectroscopy. We created preliminary
$i$ band calibrated imaging and catalogues using the generic
wide-field imaging pipeline in the Astro-WISE\footnote{Astronomical
  Wide-field Imaging System for Europe, http://www.astro-wise.org.}
survey handling system \citep{mcf13}, as operated at
OmegaCEN\footnote{OmegaCEN expertise centre for wide-field imaging,
  http://www.astro.rug.nl/~omegacen.}. The processing steps applied to
the KiDS $i$ band data involved:
\begin{itemize}
\item removing instrumental fingerprint: de-biasing, flatfielding, preliminary
  illumination correction and de-fringing,
\item primitive flagging/masking for pixels affected by saturation, cosmic
  rays and satellite tracks,
\item preliminary photometric calibration (nightly zeropoints) and astrometric
  calibration (per-chip in a single exposure),
\item swarping and coaditions of the five dithers.
\end{itemize}

The source extraction was performed using {\sc SEXTRACTOR}
\citep{ber96} configured for point source photometry. The KiDS
pointings have overlap with the first public release of the
Kilo-Degree Survey \citep{dej15}. However, the calibration used here
predates the calibration of the public release.

\begin{table}
\caption{KiDS and VIKING filters, observing time, nominal depth and typical
  seeing conditions.}
\label{tab:exptimes}
\centering
\begin{tabular}{lrrcc}
\hline \hline
Filter & $\lambda_{\rm c}$ & Exposure time & Mag limit & Seeing \\
  & (\AA) & (s) & (AB 5$\sigma$ 2 arcsec) & (arcsec) \\
\hline
$u$ & 3550 & 1000 & 24.2  & 1.0 \\
$g$ & 4775 & 900 & 25.1  & 0.8 \\
$r$ & 6230 & 1800 & 24.9 & 0.7 \\
$i$ & 7630 & 1200 & 23.5 & 0.8 \\
$Z_V$ & 8770 &  480 & 22.7 & 1.0 \\
$Y$ & 10200 & 400 & 22.0 & 1.0 \\
$J$ & 12520 & 400 & 21.8 & 0.9 \\
$H$ & 16450 & 300 & 21.1 & 1.0 \\
$K_s$ & 21470 & 480 & 21.2 & 0.9 \\
\hline
\end{tabular}
\end{table}

For VIKING data we used the images and catalogues processed with the VISTA Data
Flow System \citep{lew05,lew10} by the Cambridge Survey Unit (CASU). Catalogues
were also retrieved from the VISTA Science Archive of the Wide-field Astronomy
Unit. At the time of our analysis, we could make use of VIKING data produced
by the v1.1/1.2 CASU-VIKING pipeline ('CASUVERS=1.1/1.2'). We refer the reader
to \citet{ven13} for a description of VIKING and the data products used for
the selection of quasar candidates.

\section{Photometric selection for 5.8$<${\it z}$<$6.5 quasar candidates 
  and follow-up observations}

\label{s:QSOcandidates}

\begin{figure}
\includegraphics[width=\columnwidth]{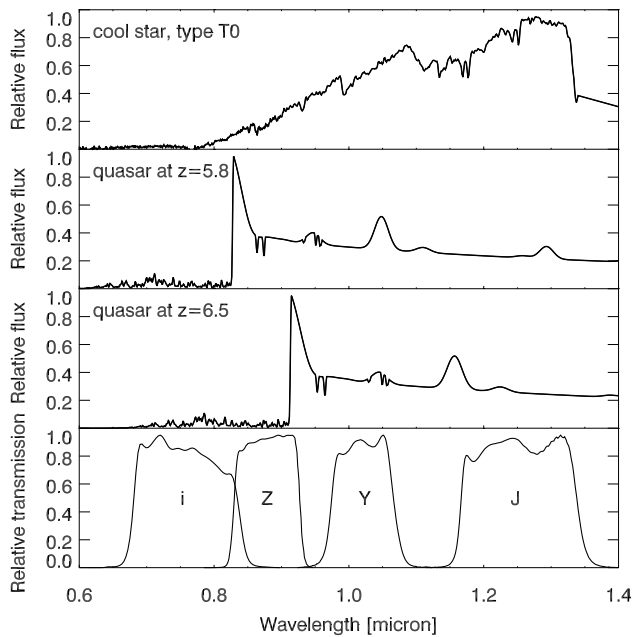}
\caption{The VIKING and KiDS filters used for the quasar search
  presented here plotted together with a (model) high-redshift quasar
  at $z=5.8$, a quasar at $z=6.5$ and one of the main contaminants, a
  cool late-type star (data taken from \citealt{geb02}). Above a
  redshift of $z\sim5.8$, the Lyman $\alpha$ line of the quasar shifts
  out of the $i$ band, which results in a large measured $i-Z_V$
  colour. Above approximately $z\sim6.5$ the Lyman $\alpha$ line falls
  outside the $Z_V$ band filter. Although both high-redshift quasars
  and cool stars are very faint in the optical $i$ band, the cool
  stars generally have a much redder $Y-J$ colour as compared to
  quasars.}
\label{z58}
\end{figure}

\subsection{Candidate selection}
\label{s:selection}

The selection of candidate quasars at $z\ga5.8$ was done as
follows. First, the KiDS $i$ band catalogues were matched with the
VIKING $Z_V$$Y$$J$$H$$K_s$ catalogues. VIKING sources that had no
match in the KiDS catalogues within a 2 arcsec radius were marked as
undetected in the \textit{i} band and a $3\sigma$ limiting magnitude
appropriate for that corresponding KiDS image was set. The median
$3\sigma$ limiting magnitude of the KiDS data we analysed was
24.0. Redshift $\sim 6$ quasar candidates were selected in the
$Z_V$ band with a signal-to-noise ratio S/N $>$ 7 and were required to
be point sources. Point sources were identified as objects with a
galaxy probability $P_\mathrm{gal}<0.95$ (see \citealt{ven13} for a
discussion on the efficiency to select point source with
$P_\mathrm{gal}<0.95$). To avoid spurious sources with only a
detection in the $Z_V$ band, we further required that an object was
also present in the $Y$ band catalogue. We subsequently applied the
following colour criteria to select quasars in the redshift range
$5.8<z<6.4$:

\smallskip

$i-Z_V>2.2$, $Z_V-Y<1.0$ and $-0.5<Y-J<0.5$. 

\smallskip

\noindent
These colour criteria are based on
the work by e.g.\ \citet{fan01} and \citet{ven07b} and are
illustrated in Fig.~\ref{z58}. In short, quasars at $z\ga5.8$ can be
identified as objects that drop-out in the $i$ band. The constraint on the
$Y-J$ colour selects against cool, red stars that can be very faint in $i$. 

The selection criteria were applied to data obtained in three areas
NGP, SGP and GAMA09. The NGP and GAMA09 area are also covered by the
SDSS. Where SDSS data were available, we applied additional colour
criteria, based on the SDSS u ($u_{\mathrm s}$), SDSS g ($g_{\mathrm
  s}$) and SDSS i ($i_{\mathrm s}$) catalogues. These additional
criteria are:

\smallskip

\noindent
($i_{\mathrm s}-Z_V>2.2$ OR undetected in $i_{\mathrm s}$) and undetected in $u_{\mathrm s}$, $g_{\mathrm s}$,

\smallskip

\noindent
where undetected was defined as having a measured magnitude fainter
than the 3$\sigma$ limiting magnitude in that band
(mag\,$>$\,mag$_\mathrm{3\sigma}$).

Due to the incompleteness of the KiDS $i$ band catalogues, especially
at fainter magnitudes, and to objects close to the edge of the KiDS
images, initially our selection criteria returned an unrealistically
large number (several thousands) of potential $z\sim6$ quasars. In the
NGP and GAMA09 region, the inclusion of the SDSS data significantly
reduced the number of candidates, especially the number of candidates
that were close to the edge of KiDS images. For remaining candidates
in those areas and those in the SGP area postage stamps in the
$i$-band were made in order to reaffirm the magnitude of the sources.

The postage stamps were analysed by measuring the flux of the
candidates in an aperture centred on the VIKING position of the
candidates. If a flux was measured with an S/N $>$ 3, and the
resulting colour was $i-Z_V<2.2$ the object was rejected as
high-redshift quasar candidate. The remaining objects were visually
inspected to confirm the reality of the candidate. A fraction was
located at the edge, or just outside the KiDS imaging area and was
removed from the list. We discovered several types of sources that
mimic the colours of high-redshift quasars in our setup. Objects
moving between the epoch of the VIKING observations and KiDS
observations and highly variable objects can have colours that are
both blue in the near-infrared bands and very red in $i-Z_V$. We were
able to remove the majority of these objects as quasar candidates by
carefully looking at the VIKING data, which are normally taken in
different nights. Images in $Z_V$, $Y$ and $J$ are generally taken in
grey time, while images in $J$, $H$ and $K_s$ are taken during bright
time. By comparing the two different $J$ band epochs and rejecting
objects with extremely blue $Y-K_s$ colours ($Y-K_s<-0.2$, while
quasars at $5.8<z<6.5$ have $0.2<Y-K_s<0.5$, e.g.\ \citealt{hew06}) we
could remove most of the moving and variable objects from our list.

In total, our final candidate list contained 30 sources: 6 objects in
the SGP region, 7 objects in the GAMA09 region and 17 candidates in
the NGP region.

\subsection{Follow-up optical imaging}
\label{s:imaging}

\begin{table}
  \caption{Followup photometry of 29 candidate $z\sim6$ quasars in the
    KiDS+VIKING catalogues. All magnitudes are in AB and limiting
    magnitudes are 3$\sigma$ limits. \label{tab:nttphot}}  \centering
\begin{tabular}{cccc}
\hline \hline
Object name & Filter & Magnitude & Telescope \\
\hline
J020935.59--323206.5 & $I_{\mathrm N}$ & $>$24.62 & NTT \\
{} & $Z_{\mathrm N}$ & $>$22.83 & NTT\\
J032835.51--325322.8 & $I_{\mathrm N}$ & 21.14$\pm$0.02 & NTT \\
J083635.31--000347.9 & $i$ & 24.36$\pm$0.45 & WHT \\
{} & $z$ & 22.51$\pm$0.21 & WHT \\
J084810.50+021948.2 & $i$ & 23.08$\pm$0.10 & WHT \\
{} & $z$ & 22.94$\pm$0.28 & WHT \\
 J090647.34--004216.0 & $I_{\mathrm N}$ & 23.46$\pm$0.43 & NTT \\
 J091026.91--011137.2 & $I_{\mathrm N}$ & 22.24$\pm$0.03 & NTT \\
 J091131.58--001305.5 & $I_{\mathrm N}$ & 23.08$\pm$0.10 & NTT \\
 J092251.90+003245.3 & $I_{\mathrm N}$ & 22.75$\pm$0.05 & NTT \\
 J113549.91+015924.4 & $I_{\mathrm N}$ & 23.07$\pm$0.07 & NTT \\
 J114833.18+005642.3 & $I_{\mathrm N}$ & 22.67$\pm$0.05 & NTT \\
 {} & $Z_{\mathrm N}$ & 21.46$\pm$0.15 & NTT \\
 J115049.81--013830.2 & $I_{\mathrm N}$ & 23.20$\pm$0.06 & NTT \\
 J121144.38+005348.8 & $I_{\mathrm N}$ & 23.46$\pm$0.08 & NTT \\
J121404.53+020220.2 & $I_{\mathrm N}$ & 22.38$\pm$0.10 & NTT \\
J121516.88+002324.7 & $I_{\mathrm N}$ & 22.93$\pm$0.10 & NTT \\
 J122839.49+010253.7 & $I_{\mathrm N}$ & 22.13$\pm$0.02 & NTT \\
 J140552.33+030110.3 & $I_{\mathrm N}$ & 22.65$\pm$0.04 & NTT \\
 J141157.27--001613.5 & $I_{\mathrm N}$ & 22.92$\pm$0.12 & NTT \\
 J142054.44+021247.3 & $I_{\mathrm N}$ & 21.88$\pm$0.02 & NTT \\
 J142300.93+011240.0 & $I_{\mathrm N}$ & 22.53$\pm$0.09 & NTT \\
 J142426.55+011304.1 & $I_{\mathrm N}$ & 22.75$\pm$0.04 & NTT \\
 J143428.30--010043.1 & $I_{\mathrm N}$ & 22.69$\pm$0.03 & NTT \\
 J144130.49+004119.7 & $I_{\mathrm N}$ & 22.52$\pm$0.03 & NTT \\
 J144922.61--004404.9 & $I_{\mathrm N}$ & 22.85$\pm$0.04 & NTT \\
 J145250.27--004047.8 & $I_{\mathrm N}$ & 23.22$\pm$0.05 & NTT \\
 J145442.58--020118.8 & $I_{\mathrm N}$ & 22.59$\pm$0.03 & NTT \\
J224319.75--331506.8 & $I_{\mathrm N}$ & $>$23.70 & NTT \\
{} & $Z_{\mathrm N}$ & $>$23.24 & NTT \\
J225656.49--333910.1 & $I_{\mathrm N}$ & 22.10$\pm$0.05 & NTT \\
J232355.24--303801.7 & $I_{\mathrm N}$ & 22.82$\pm$0.05 & NTT \\
{} & $Z_{\mathrm N}$ & 22.07$\pm$ 0.12 & NTT \\
J234331.78--313627.5 & $I_{\mathrm N}$ & 21.30$\pm$ 0.03 & NTT \\
\hline
\end{tabular}
\end{table}

We took images of 29 of our 30 $z\sim6$ quasar candidates during four
separate observing runs. The first run was on 2012 June 29--30 at the
3.58 m ESO New Technology Telescope (NTT) using the European Southern
Observatory's Faint Object Spectrograph and Camera 2
\citep[EFOSC2;][]{buz84}. We observed in the filters Gunn $i$ (filter
$i$ no.\ 705, hereafter $I_{\mathrm N}$) and Gunn $z$ (filter $z$ no.\ 623,
hereafter $Z_{\mathrm N}$). The second observing run was also with
NTT/EFOSC2 on 2012 July 23--25. Thirdly, two candidates were observed
with ACAM \citep{ben08} on the William Herschel Telescope (WHT) on
2012 November 10 in service mode, using the filters $i$ and
$z$. Finally, we used the NTT again on 2013 March 13--15 to follow-up
candidates with EFOSC2.

The candidates were observed with total exposure times between 720
and 2880\,s. The integration time for each individual target depended
on the sky transparency, seeing and expected brightness of the
source. Data reduction steps included bias subtraction, flat fielding
using sky flats and sky subtraction using the unregistered science
frames. The images were calibrated using bright, unsaturated point
sources within the field of view. Since the ACAM $i$ filter is very
similar to the VST $i$ band, the zero-point of the ACAM $i$ images
were directly derived from the KiDS magnitudes of point sources in the
field. The zero-points of the NTT images were computed using relations
between $I_{\mathrm N}$ and $Z_{\mathrm N}$ and SDSS and VIKING bands
derived from synthesized colours of stars \citep{ven13}. The results
of the follow-up photometry are listed in Table~\ref{tab:nttphot}.

\begin{figure}
\centering
\includegraphics[width=\columnwidth]{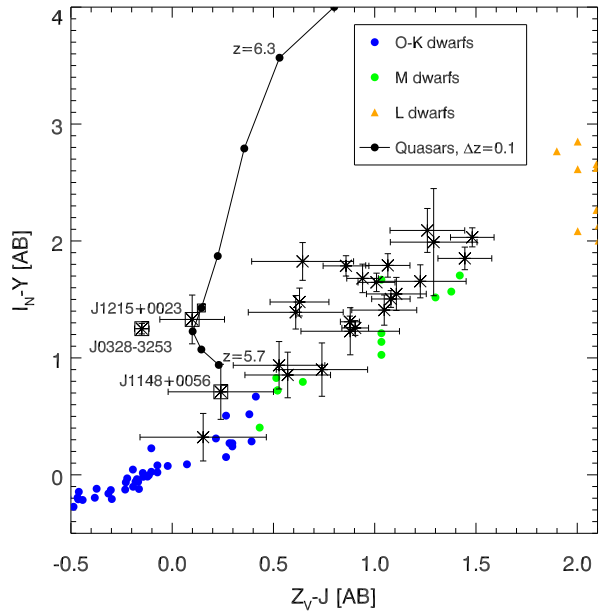}
\caption{$I_{\mathrm N}-Y$ versus $Z_V-J$ diagram illustrating the
  outcome of the follow-up NTT imaging of our $z\sim6$ quasar
  candidates. The blue and green circles represent the colours of main
  sequence stars and M dwarfs, respectively. The yellow triangles are
  simulated colours of L dwarfs, and the connected black circles
  illustrate the colours of quasars at various redshifts $z>5.7$. The
  crosses with error bars represent the candidates observed with the
  NTT. Based on the $I_{\mathrm N}-Z_V$, $I_{\mathrm N}-Z_{\mathrm
    N}$, $I_{\mathrm N}-Y$ and $Z_V-J$ colours most observed
  candidates were identified as foreground interlopers. The open
  squares are three objects that remained good candidates after the
  follow-up imaging and for which we obtained optical spectroscopy
  (see Section~\ref{s:spectroscopy}). Additional $Z_{\mathrm N}$
  photometry (see Table~\ref{tab:nttphot}, not shown in this figure)
  confirmed that J1148+0056 was a good high-redshift quasar
  candidate. Note that for a fourth good quasar candidate, J0839+0015,
  we did not obtain follow-up photometry (see Section~\ref{s:j0839}).}
\label{ccfollowup}
\end{figure}

The follow-up imaging rejected 26 of the 29 observed targets as
potential $z\sim6$ quasars. The majority of rejected candidates have
colours consistent with M dwarfs, see Fig.~\ref{ccfollowup}. The four
objects that remained good quasar candidates were J083955.36+001554.2
(hereafter J0839+0015), J114833.18+005642.3 (hereafter J1148+0056),
J121516.88+002324.7 (hereafter J1215+0023) and J032835.51--325322.8
(hereafter J0328--3253). Note that for J0839+0015 we did not obtain
any follow-up photometry (see Section~\ref{s:j0839}). For the four
remaining objects we took an optical spectrum to confirm their nature.

\subsection{Spectroscopic observations}
\label{s:spectroscopy}

\begin{figure*}
\centering
\includegraphics[width=\textwidth]{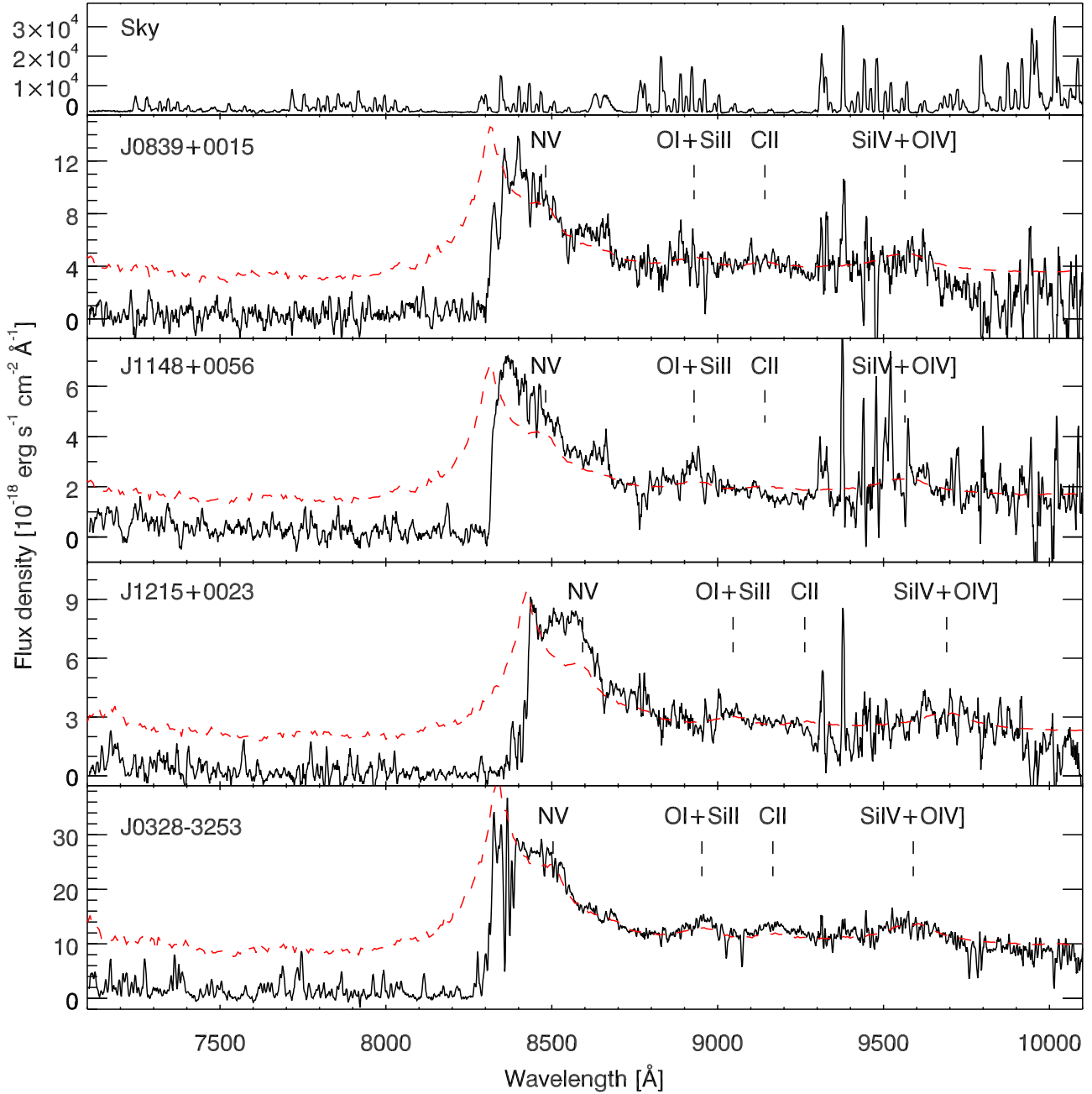}
\caption{Discovery spectra from VLT/FORS2 of three $z\sim6$ quasars in the
  KiDS+VIKING survey. The spectra of J0839+0015 and J1215+0023 are boxcar
  averaged over five pixels. For comparison, a spectrum of the sky is plotted
  at the top. The position of several emission lines are marked. The spectra
  are overplotted with the composite quasar spectrum from \citet{van01}.}
\label{spectra}
\end{figure*}

The spectroscopic observations of our four remaining quasar candidates were
carried out with the FOcal Reducer/low dispersion Spectrograph 2
\citep[FORS2;][]{app98} on the 8.2 m Very Large Telescope (VLT) Antu. For all
observations we used a long slit with a width of 1.3\,arcsec and the grism
`600z' with a central wavelength of 9010\,\AA\ and a resolution of
$R=1390$. J0328--3253 was observed on 2012 October 7 for 1782\,s in clear
conditions with a seeing around 0.8\,arcsec. On 2012 December 12 J0839+0015 was
observed for 1782\,s with a seeing ranging between 1.2\,arcsec and
1.7\,arcsec. J1215+0023 was observed on 2013 February 7 and 2013 March 5 for a
total integration time of 5280\,s. The seeing during the February observations
varied between 1.4\,arcsec and 1.7\,arcsec, while in March the conditions were
better with a seeing around 0.7\,arcsec. Finally, a spectrum of J1148+0056 was
taken on 2013 April 13 for 2697\,s in a seeing of 0.9\,arcsec. All observations
were setup in such a way that a nearby bright star was included in the slit to
provide a trace for the candidate spectrum.

Daytime exposures of He, Ar and Ne lamps were obtained to provide the
wavelength calibration. By measuring the wavelength of sky emission
lines in our science spectra we determined that the wavelength
calibration had a typical uncertainties between 0.2 and 0.5\,\AA. For
the flux calibration we used observations with a 5\,arcsec long slit
(taken the same night as the science observations) of one of the
spectrophotometric standard stars Feige 110, HD49798, LTT4816 and
LTT7379 \citep{oke90,ham92,ham94}. Finally the flux calibrated spectra
were scaled to the VIKING $Z_V$ band magnitudes to account for slit
losses.

The reduced spectra of the four quasar candidates are shown in
Fig.~\ref{spectra}.

\section{Four quasars around a redshift of 6}
\label{s:newqsos}

The four targets of our follow-up spectroscopy all show a blue
continuum with a sharp break between 8300 and 8500\,\AA,
characteristic of quasars at $z\sim6$. To measure the redshift of the
newly identified quasars, we measured the redshifted position of
emission lines in the spectra. When visible, the emission lines we
used for this purpose are O\,{\sevensize I} + Si\,{\sevensize II} at
$\lambda_\mathrm{rest}=1305.42$\,\AA, C\,{\sevensize II} at
$\lambda_\mathrm{rest}=1336.6$\,\AA\ and Si\,{\sevensize IV} +
O\,{\sevensize IV}] at $\lambda_\mathrm{rest}=1398.33$\,\AA. The
  central wavelength of the emission lines are taken from the
  composite quasar spectrum by \citet{van01}. To compute the absolute
  magnitudes of the quasars at a rest-frame wavelength of
  1450\,\AA\ ($M_{1450}$), we fitted a power-law slope in regions
  without emission lines. The regions used were
  $1270<\lambda_\mathrm{rest}$/\AA$<1295$,
  $1315<\lambda_\mathrm{rest}$/\AA$<1325$ and
  $1345<\lambda_\mathrm{rest}$/\AA$<1370$. Below we describe the new
  quasars in more detail. A summary of the properties of the quasars
  is given in Table~\ref{tab:newqsos}.

\subsection{J0328$-$3253 at \boldmath{$z=5.86$}}

This quasar was selected as a high signal-to-noise ratio candidate in
our KiDS+VIKING catalogues. This source has a $Z_V$ band magnitude of
$Z_V=19.8$ and blue infrared colours of $Z_V-Y=-0.1$ and $Y-J=-0.1$
(Table~\ref{tab:newqsos}). The source was detected in the $i$ band
with a magnitude of $i=22.14\pm0.06$ and thus had a
$i-Z_V=2.3$. Follow-up photometry with the NTT confirmed a break in
the spectral energy distribution of the source shortwards of the
$Z_V$ band. The $I_{\mathrm N}-Z_V=1.3$, consistent with a quasar at
$z\sim5.8$. VLT/FORS2 spectroscopy confirmed that the source is at
$z=5.86\pm0.03$ (Fig.~\ref{spectra}). We measured the redshift using
the emission lines O\,{\sevensize I} + Si\,{\sevensize II},
C\,{\sevensize II} and Si\,{\sevensize IV} + O\,{\sevensize IV}] and
  found a redshift around $z=5.86$ for all three lines. Our power-law
  fit to the continuum gives an absolute magnitude of
  $M_{1450}=-26.60\pm0.04$. After our discovery of this quasar in
  2012, the object was independently selected as a bright quasar
  candidate in the VST-ATLAS survey \citep{carn15}.

\subsection{J0839+0015 at \boldmath{$z=5.84$}}
\label{s:j0839}

This object was initially selected with $i-Z_V>2.7$, $Z_V-Y=-0.1$ and
$Y-J=0.3$ (Table~\ref{tab:newqsos}). Analysis of the $i$ band postage
stamp revealed a faint source with a magnitude of $i=23.46\pm0.33$,
resulting in an $i-Z_V\sim2.4$. The two epochs of near-infrared data
did not show strong variability and the detection of a source in the
$i$ band ruled out that this object was moving. We therefore decided
to take a spectrum of this source. Despite poor observing conditions
the resulting spectrum clearly shows a break at
$\sim$8300\,\AA\ (Fig.~\ref{spectra}). If this break is at the
wavelength of Lyman $\alpha$, then the redshift would be around
$z\sim5.83$. This redshift was confirmed by measuring the position of
faint emission lines. The O\,{\sevensize I} + Si\,{\sevensize II} gave
a redshift of $z\sim5.81$, while the Si\,{\sevensize IV} +
O\,{\sevensize IV}] line gave $z\sim5.87$. We therefore estimate that
  the redshift of this quasar is $z=5.84\pm0.04$. The continuum fit
  gives an absolute magnitude of $M_{1450}=-25.36\pm0.11$.

\subsection{J1148+0056 at \boldmath{$z=5.84$}}

The source J1148+0056 was selected as a faint ($Z_V=21.8$) candidate
with colours $i-Z_V>2.3$, $Z_V-Y=-0.2$ and $Y-J=0.4$. The $i$ band
postage stamp showed no source at the position of the $Z_V$ band
object. Follow-up imaging with the NTT measured a colour of
$I_{\mathrm N}-Z_{\mathrm N}=1.2$ and $Z_{\mathrm N}-J=-0.3$,
consistent with a high redshift quasar. Optical spectroscopy detected
a source with continuum longwards of $\lambda\sim8300$\,\AA\ and faint
emission lines at 8930 \AA\ and at 9115\,\AA. We identified these
lines with O\,{\sevensize I} + Si\,{\sevensize II} and C\,{\sevensize
  II} redshifted to $z=5.84\pm0.03$. From a power-law fit to the
continuum we estimate that the absolute magnitude of this quasar is
$M_{1450}=-24.46\pm0.11$, which makes it the faintest quasar in our
sample.

\subsection{J1215+0023 at \boldmath{$z=5.93$}}

J1215+0023 was a high priority quasar candidate in our lists with
colours $i-Z_V>2.7$, $Z_V-Y=-0.2$ and $Y-J=0.3$. NTT imaging confirmed
that the source had a break in the spectral energy distribution
shortwards of the $Z_V$ band ($I_{\mathrm N}-Z_V=1.5$). From the FORS2
spectrum we measure a redshift of $z=5.93\pm0.03$, based on the
O\,{\sevensize I} + Si\,{\sevensize II} and Si\,{\sevensize IV} +
O\,{\sevensize IV}] lines. From the continuum we measure an absolute
  magnitude of $M_{1450}=-24.67\pm0.14$.

\begin{table*}
\caption{Properties of the newly discovered quasars. The redshift uncertainty
  does not include the uncertainty between the redshift of the UV emission lines
  and the systemic redshift of the quasars. The uncertainty in the absolute
  magnitude is a combinations of the uncertainty in the continuum fit and in
  the absolute scaling of the spectrum based on the $Z_V$-band magnitude.}
\label{tab:newqsos}
\centering
\begin{tabular}{ccccccccc}
  \hline \hline
  Object name & RA (J2000) & Dec. (J2000) & Redshift & 
  $Z_\mathrm{VIKING}$ & $Y_\mathrm{VIKING}$ & $J_\mathrm{VIKING}$ &
  $K_{s,\mathrm{VIKING}}$ & $M_{1450}$ \\
  \hline
  J0328$-$3253 & 03$^{\mathrm h}$28$^{\mathrm m}$35\fs511 &
  --32$\degr$53\arcmin22\farcs92 & 5.86$\pm$0.03 & 19.83$\pm$0.02 &
  19.89$\pm$0.04 & 19.98$\pm$0.03 & 19.69$\pm$0.05 & --26.60$\pm$0.04
  \\
  J0839+0015 & 08$^{\mathrm h}$39$^{\mathrm m}$55\fs356 &
  +00$\degr$15\arcmin54\farcs21 & 5.84$\pm$0.04 & 21.09$\pm$0.05 &
  21.19$\pm$0.10 & 20.89$\pm$0.11 & 20.50$\pm$0.15 & --25.36$\pm$0.11
  \\
  J1148+0056 & 11$^{\mathrm h}$48$^{\mathrm m}$33\fs180 &
  +00$\degr$56\arcmin42\farcs26 & 5.84$\pm$0.03 & 21.79$\pm$0.10 &
  21.96$\pm$0.23 & 21.55$\pm$0.24 & 21.20$\pm$0.23 & --24.46$\pm$0.11
  \\
  J1215+0023 & 12$^{\mathrm h}$15$^{\mathrm m}$16\fs879 &
  +00$\degr$23\arcmin24\farcs74 & 5.93$\pm$0.03 & 21.42$\pm$0.07 &
  21.60$\pm$0.18 & 21.32$\pm$0.14 & 21.10$\pm$0.21 & --24.67$\pm$0.14
  \\ \hline
\end{tabular}
\end{table*}

\section{Discussion and outlook}
\label{s:summary}

\begin{figure}
\centering
\includegraphics[width=\columnwidth]{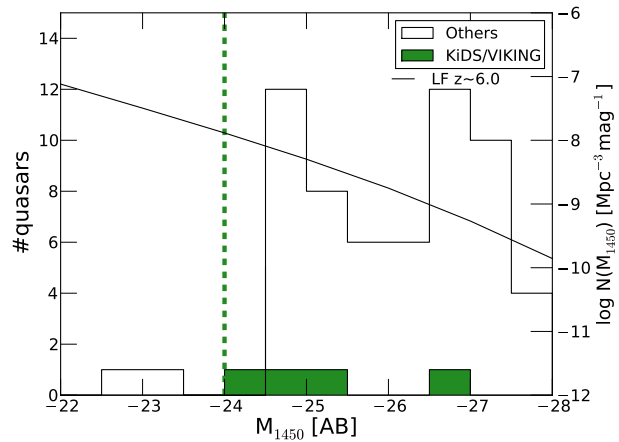}
\caption{Distribution of quasars in the redshift range $5.8<z<6.4$ as
  function of absolute magnitude. Plotted are the histograms of the 60
  previously published quasars in this redshift range (white) and new
  quasars discovered with the KiDS/VIKING survey pilot (green). The
  solid black curve indicates the luminosity function for quasars at
  redshift 6 as derived by \citet{kas15}, using their `case 1'. The
  dashed green line indicates the selection limit, i.e., the
  intrinsically faintest quasars that we could select with the
  KiDS/VIKING survey. The aim of the KiDS/VIKING survey is to triple
  the number of known quasars fainter than $M_\mathrm{1450}>-25$.}
\label{f:lumfunct}
\end{figure}

By combining KiDS $i$ band data with VIKING $Z_V, Y, J, H, K_s$ catalogues,
we performed a pilot study to search for faint $z\sim6$ quasar
candidates. In our first year of obtaining follow-up data, we
discovered four new quasars in an area of $\sim$254 deg$^2$. To compare
our quasars with those of other high redshift quasar searches we plot
in Fig.~\ref{f:lumfunct} the $z\sim6$ quasar luminosity function from
\citet{kas15} as function of absolute magnitude. Overplotted is a
histogram of the 60 known quasars at $5.8<z<6.4$, taken from the literature
\citep[][and references
  therein]{zei11,ban14,cal14,kas15,carn15,jia15,ree15}. The new
quasars from the KiDS/VIKING surveys are generally fainter ($Z_V$ mag
$\ga$ 21) than the quasars found in, for example, the SDSS and PS1
surveys, showing that our $z\sim6$ quasar search is probing the
faint end of the known $z=6$ quasar population.

One of the questions is how many quasars we were expecting in our
survey. To estimate this, we took the $z\sim6$ quasar luminosity
function and computed the expected number of quasars as function of
area and limiting magnitude. The limiting magnitude varied from area
to area, and mostly depended on the depth of the KiDS $i$-band
imaging. The median depth to which we could search for quasars was
$Z_V\sim21.65$. We integrated the \citet{wil10a} luminosity
function\footnote{Using the \citet{kas15} luminosity function instead
  of the \citet{wil10a} luminosity function gives the same result.}
and in the redshift range $5.8<z<6.4$ we expected $\sim$$7$ quasars in
the area we surveyed, where we discovered four. The difference between
the predicted number and the observed number of quasars might well be
due to various sources of incompleteness of our search. The most
prominent ones are: catalogue completeness, point source selection
efficiency and completeness of the colour selection. In \citet{ven13}
it was found that the VIKING catalogue completeness is $\sim$96\%
above a S/N of 7, while the fraction of point sources that is
incorrectly classified as extended objects is around 6\%--7\%. The
largest incompleteness is the colour incompleteness: both due to
random photometric errors and due to small variations in the quasar
spectral energy distribution and in the intergalactic medium (IGM)
absorption, quasars can be placed outside our colour selection
box. For example, quasars with a redshift around $z\sim5.8$ have an
$i-Z_V$ colour that is a steep function of redshift, see e.g.\ fig.~1
in \citet{ban14}. It is therefore possible that we are missing quasars
around $z\sim5.8$ that have $i-Z_V<2.2$. Indeed, by matching our data
with high-redshift quasars that have previously been discovered, we
found that the SDSS quasar J0836+0054 at $z=5.82$ \citep{fan01} is
detected in both the VIKING and KiDS survey. While the redshift of
this quasar is very close to the ones we present in this paper,
J0836+0054 has an $i-Z_V=1.70$, well below our cut of $i-Z_V=2.2$. At
the high end of our probed redshift range, $z\sim6.4$, due to
variations in the IGM and in the quasar's emission line strength and
due to photometric uncertainties, quasars can become too red in
$Z_V-Y$ to be selected by our colour criterion ($Z_V-Y<1.0$, see
Section~\ref{s:selection}). In \citet{ven13} it was shown through
modelling that up to 15\% of artificial quasars at $z=6.4$ fulfilled
the colour selection $Z_V-Y>1.1$. This indicates that we are also
incomplete in our selection of quasars at $z=6.3-6.4$. Combined with
the other sources of incompleteness, the discovery of four new quasars is
well in line with the theoretical expectation of seven.

The total area that ultimately will be covered by the KiDS and VIKING
surveys is 1500 deg$^2$, roughly six times the area we searched in our
pilot study. We therefore expect to discover at least 20 additional
quasars in this area. Furthermore, there are several ways to push our
quasar search to even fainter magnitudes. For example, when selecting
quasar candidates, we could reduce the $i-Z_V$ colour limit to
$i-Z_V=2.0$ (similar to the PS1 quasar search, see
\citealt{ban14}). This will increase our completeness, especially at
the low-redshift end of our quasar search, $z\sim5.8$. Additionally,
our requirement for a candidate to have a detection in the $Y$ band,
while significantly reducing the number of spurious sources, limits in
certain areas how faint we can select candidates as the $i$  and
$Z_V$ band catalogues are deeper than the $Y$-band catalogue (see
Table~\ref{tab:exptimes}). By pushing our selection to fainter
magnitudes, we should be able to uncover quasars as faint as
$M_\mathrm{1450}=-24$. Based on the quasar luminosity function, we
therefore expect to discover at least 30 more quasars down to
$M_{1450}=-24$, significantly increasing the number of quasars at
these faint absolute magnitudes.

Besides exploring the new survey data that will become available, our
plan is to characterize the newly discovered quasars with
near-infrared, far-infrared and sub-mm spectroscopy. This will allow
us to constrain black hole masses, ISM metallicity and host
masses. This is important to understand the relationship between the
accreting black hole and the stellar component of galaxies in the
first billion years of the Universe.

\section*{Acknowledgements}
  
The authors are grateful for advice and support by the KiDS Production
Team during this early pilot programme in the first year of KiDS
operations. This work is financially supported by the Netherlands
Research School for Astronomy (NOVA) and Target. Target is supported
by Samenwerkingsverband Noord Nederland, European fund for regional
development, Dutch Ministry of economic affairs, Pieken in de Delta,
Provinces of Groningen and Drenthe. Target operates under the auspices
of Sensor Universe. BPV acknowledges funding through the ERC grant
`Cosmic Dawn'. EB thanks the IMPRS for Astronomy \& Cosmic Physics
at the University of Heidelberg. This work is supported by the
Netherlands Organization for Scientific Research (NWO) through grant
614.061.610 (JdJ). This publication has made use of data from the
VIKING survey from VISTA and KiDS survey from the VST at the ESO
Paranal Observatory, under programme IDs 179.A-2004, 177.A-3016,
177.A-3017 and 177.A-3018. Data processing has been contributed by the
VISTA Data Flow System at CASU, Cambridge and WFAU, Edinburgh and the
Astro-WISE Data Flow System at OmegaCEN, Kapteyn Astronomical
Institute, University of Groningen. {\it Author Contributions}: all
authors contributed to the development and writing of this paper. The
authorship list reflects the lead authors (BV, GVK, JM, EV) followed
by two alphabetical groups. The first alphabetical group (EB, RD)
includes those who are key contributors to both the scientific
analysis and the data products. The second group covers those who have
either made a significant contribution to the data products, or to the
scientific analysis.

\label{lastpage}

\end{document}